\documentclass[3p]{elsarticle}

\usepackage{rotating}
\usepackage{graphicx}
\usepackage{amssymb}
\usepackage{amsmath}
\usepackage{tabls}
\usepackage{boxedminipage}
\usepackage{color}
\usepackage{bm}
\usepackage{bbm}
\usepackage{maybemath}

\newcommand{\dd}{\mathrm{d}}

\begin{document}

\title{Sources, Potentials and Fields in Lorenz and Coulomb Gauge:\\
Cancellation of Instantaneous Interactions for Moving Point Charges}

\author{B. J. Wundt}
\author{U. D. Jentschura$^*$}

\address{Missouri University of Science and Technology,
Physics Department, Rolla, Missouri 65409--0640, USA\\[0.33ex]
$^*${\rm{email: ulj@mst.edu, telephone +1-573-3416221, 
telefax +1-314-3414715}}}

\begin{abstract}
We investigate the coupling of the electromagnetic 
sources (charge and current densities) to the 
scalar and vector potentials in classical electrodynamics,
using Green function techniques.
As is well known, the scalar potential shows
an action-at-a-distance behavior in Coulomb gauge.
The conundrum generated by the instantaneous 
interaction has intrigued physicists for a long time.
Starting from the differential equations 
that couple the sources to the potentials, we here show
in a concise derivation, using the 
retarded Green function, 
how the instantaneous interaction cancels in the 
calculation of the electric field.
The time derivative of a specific additional term 
in the vector potential, present only in Coulomb gauge,
yields a supplementary contribution to the electric field
which cancels the gradient of the  
instantaneous Coulomb gauge scalar potential,
as required by gauge invariance.
This completely eliminates the contribution of the instantaneous interaction
from the electric field.
It turns out that a careful formulation of the
retarded Green function, inspired by field theory,
is required in order to correctly treat boundary terms in 
partial integrations.
Finally, compact integral representations are derived 
for the Li\'{e}nard--Wiechert potentials (scalar and vector) in 
Coulomb gauge which manifestly contain two compensating 
action-at-a-distance terms.
\end{abstract}


\maketitle


%
%
\section{Introduction}

The instantaneous coupling~\cite{Ja1998} of the electrostatic potential to the
charge density in Coulomb gauge (``action at a distance'') has given rise to a
number of concerns, recorded in the literature.  Rohrlich has addressed the
issue in Ref.~\cite{Ro2002}. He argues that the seemingly instantaneous
integral for the scalar potential can be rewritten as an integral involving the
retarded Green function, with a nonstandard source term that does not only
involve the charge density but also the longitudinal part of the current
density. Using the special form of the equations that relate the charge density
to the longitudinal part of the current density in Coulomb gauge, the retarded
integral may be shown to be equal~\cite{Ro2002} to the instantaneous integral,
thus clarifying that the instantaneous nature of the action-at-a-distance
formula is a particular feature of the Coulomb gauge that does 
not contradict the causality principle. 
The problem can be sidestepped if one formulates electromagnetic 
theory purely in terms of fields rather than potentials, 
using Dirac-like equations~\cite{Mo2010}, but one might 
otherwise argue that the gauge potentials
which enter the nontrivial phase of the wave function in the Aharonov--Bohm
effect, are assigned some kind of physical meaning within the concept of 
covariant derivatives in field theory,
even if one can otherwise express the Aharonov--Bohm phase 
in terms of the magnetic flux entering the loop~\cite{AhBo1959,JaOk2001}.

One could argue that despite Rohrlich's arguments~\cite{Ro2002},
the instantaneous form of the scalar potential still poses potential 
problems with regard to causality, even if it is possible to
rewrite it in retarded form. Namely, it could be argued that it is still not
clear how the instantaneous part of the potential actually cancels when the
electric and magnetic fields are calculated, even though it is possible to give
an alternative, retarded form for the integral defining the scalar potential.
Indeed, it would be somewhat discomforting if the action-at-a-distance part of
the interaction were to prevail in the final result for the field strength. 
In Ref.~\cite{Ja2002}, Jackson investigates this problem in the framework of 
gauge potentials and field equations.

We here attempt to provide an alternative view of the problem, 
inspired by field theory, where the retarded Green function is closely
related to the commutator of the fields [see Eq.~(2.55) of \cite{PeSc1995}]. 
We start from the structure of the equations that couple 
the sources and potentials in Coulomb gauge,
solve these using the retarded Green function, and show 
the expected cancellation of instantaneous terms in the observable
electric and magnetic fields. 
The correct and, in our case, required formulation of the retarded Green 
function in space-time coordinates reads as 
\begin{equation}
\label{GR}
G_R(\vec r, t, \vec r', t') =
\frac{c}{4\pi \, \epsilon_0} \; \frac{\Theta(t - t')}{|\vec r - \vec r'|}
\biggl[
\delta\left( |\vec r - \vec r'| - c \, (t - t')\right) -
\delta\left( |\vec r - \vec r'| + c \, (t - t')\right)
\biggr] \,,
\end{equation}
where $\Theta$ is the step function, $c$ is the speed of light,
the coordinates are $\vec r$ and $\vec r'$, and the times are 
$t$ and $t'$.  We keep {\em both} terms in square brackets 
until the final stage of the calculation. In particular, 
we notice that the expression~\eqref{GR} formally is 
at variance with Eq.~(31.45) of Schwinger's well-known
textbook~\cite{ScDRMiTs1998}, 
where the term proportional to $\Theta(t - t') \, 
\delta\left( |\vec r - \vec r'| + c \, (t - t')\right)$ is discarded.
In many cases (some physicists would say in {\em most} cases), 
it is indeed possible to discard 
said term [see Eq.~\eqref{GRsimple} below], as done in Ref.~\cite{ScDRMiTs1998}.
However, in any representation of the Dirac $\delta$ distribution,
e.g., formulated in terms of normalized Gaussians whose width tend to zero, 
the step function will assume values of unity within the 
width of the representation of the 
Dirac $\delta$ function when $|\vec r - \vec r'| \to 0^+$ 
and $c \, (t - t')\to 0^+$. {\em One thus cannot discard any term in 
Eq.~\eqref{GR} if one would like to carry out partial integrations 
correctly; the latter are useful in the investigation of the coupling of the 
sources to the potentials and fields.}

The general results derived in the current work
are then verified on the basis of a concrete problem,
namely, the potentials generated by a moving point charge. 
Due to the singular character of the source terms in this 
problem, a treatment using 
the retarded Green is seen to be more convenient. In classical
electrodynamics, all electric and magnetic fields can in principle be described as a
superposition of fields generated by infinitesimal moving point charges. 
Again, the intriguing problem is that in Coulomb
gauge (radiation gauge), the scalar potential $\Phi$ can be written as an
instantaneous integral over charges that are far away from the observation
point. A change in a charge distribution displaced by 
even astronomical distance scales
therefore leads to an instantaneous change in the scalar potential
at an observation point on Earth.  In order to address this latter question, we
here not only calculate the scalar potential, but also the vector potential
generated by a moving charge, in Coulomb gauge and in general form. This
problem has independently attracted some 
interest~\cite{La1999coulomb,Hn2004,Es2007}. 

In this article, we use SI mksA units throughout.  We start with a number of
general considerations in Sec.~\ref{gen}, before calculating the potentials in
Coulomb gauge (Sec.~\ref{pot}).  Particular emphasis is laid on the correct use
of the full space-time representation of the retarded Green function, which is
crucial for the cancellation of a few singular terms. Conclusions are reserved
for Sec.~\ref{conclu}.

%
%
\section{General Considerations}
\label{gen}

%
%
\subsection{Potentials and Sources}

In electrodynamics, we differentiate the Lorentz force and the Lorentz
transformation from the Lorenz gauge.
The Lorenz gauge is Lorentz-covariant~\cite{LorenzLorentz}.
The Lorenz condition for the scalar 
potential $\Phi$ and the vector potential $\vec A$ reads
\begin{subequations}
\begin{equation}
\vec\nabla \cdot \vec{A}_L\left( \vec{r},t\right) +
\frac{1}{c^2} \, 
\frac{\partial }{\partial t}\Phi_L\left( \vec{r},t\right) =0  \,.
\end{equation}
The subscript $L$ will be used throughout this 
article in order to denote the Lorenz gauge.
The potentials are coupled to the sources by inhomogeneous wave 
functions,
\begin{eqnarray}
\label{scalwave}
\left( \frac{1}{c^{2}}\frac{\partial^{2}}{\partial t^{2}} 
- \vec\nabla^2 \right) \; \Phi_L\left( \vec{r},t\right) 
&=& \frac{1}{\epsilon_0} \, \rho\left( \vec{r},t\right) \,,
\\[0.7ex]
\label{vectwave}
\left( \frac{1}{c^{2}}\frac{\partial^{2}}{\partial t^{2}} 
- \vec\nabla ^{2} \right) \; \vec{A}_L\left( \vec{r},t\right) 
&=& \mu_0 \, \vec{J}\left( \vec{r},t\right) \,,
\end{eqnarray}
\end{subequations}
where $\epsilon_0$ and $\mu_0$ are the vacuum permittivity 
and vacuum permeability, respectively.
The charge density is $\rho\left( \vec{r},t\right)$,
and the current density is $\vec J\left( \vec{r},t\right)$.
In Lorenz gauge, all potentials and fields 
manifestly propagate with
the speed of light in vacuum $c$. It is possible to define
gauges, where the potentials propagate with different speeds~\cite{Ja2002}.
For quantum electrodynamics, this is investigated in Ref.~\cite{Ba1991}.

In the Coulomb gauge (subscript $C$), or radiation gauge, the 
gauge condition is
\begin{subequations}
\label{rgauge}
\begin{equation}
\label{rgaugeX}
\vec\nabla \cdot \vec{A}_C\left( \vec{r}, t\right) = 0 \,,
\qquad
\vec{A}_C\left( \vec{r}, t\right) =
\vec{A}_{C\perp}\left( \vec{r}, t\right) \,,
\end{equation}
which states that the Coulomb gauge vector 
potential $\vec A$ is equal to its transverse component $\vec{A}_{\perp}$.
The coupling to the 
sources is governed by the following equations,
\begin{align}
\label{rgauge1}
\vec\nabla^2 \Phi_C\left( \vec{r}, t\right) =& \;
-\frac{1}{\epsilon_0} \rho\left( \vec{r},t\right) \,,
\\[0.7ex]
\label{rgauge2}
\left( \frac{1}{c^{2}}\frac{\partial^2}{\partial t^2} 
-\vec\nabla^{2} \right)
\vec{A}_{C\perp}\left( \vec{r}, t\right) =& \;
\mu_0 \, \vec{J}_\perp\left( \vec{r}, t\right) \,,
\\[0.7ex]
\label{rgauge3}
\epsilon_0\, \frac{\partial }{\partial t} 
\vec\nabla \Phi_C\left( \vec{r},t\right) 
=& \; \vec{J}_\parallel(\vec r, t) \,.
\end{align}
\end{subequations}
Here, we refer to the longitudinal and transverse 
components of a general vector field $\vec J\left( \vec r, t\right)$ 
as $\vec J_\parallel\left( \vec r, t\right)$ 
and $\vec J_\perp\left( \vec r, t\right)$, respectively.
According to a well-known theorem,
any vector field can be uniquely decomposed
into a longitudinal and a transverse component, as follows,
\begin{equation}
\vec J\left( \vec{r}, t\right) =
\vec{J}_\perp\left( \vec{r}, t \right) +
\vec{J}_\parallel\left( \vec{r}, t \right) \,,
\qquad
\vec \nabla \cdot \vec{J}_\perp\left( \vec{r}, t\right) = 0 \,,
\qquad 
\vec \nabla \times \vec{J}_\parallel\left( \vec{r}, t\right) = 0 \,.
\end{equation} 
Given $\vec J\left( \vec{r}, t\right)$, the 
parallel and longitudinal components can be computed as 
\begin{subequations}
\label{longtran}
\begin{align}
\label{long}
\vec{J}_\parallel\left( \vec{r},t\right) =& \;
-\frac{1}{4\pi } \vec\nabla 
\int \frac{\vec\nabla' \cdot \vec{J}\left( \vec r', t\right) }%
{\left| \vec r- \vec r' \right| }\,\dd^{3} r' \,, \\
\label{tran}
\vec{J}_\perp\left( \vec{r},t\right) =& \;
\frac{1}{4\pi} \vec\nabla \times
\int \frac{\vec\nabla' \times \vec{J}\left( \vec r',t\right) }%
{\left| \vec r- \vec r' \right| }\,\dd^{3}r^{\prime } \,.
\end{align}
\end{subequations}
These are highly non-local functions of the 
full current density $\vec{J}\left( \vec r,t\right)$,
as pointed out in Ref.~\cite{Ro2002}.

%
%
\subsection{Potentials and Fields}

The general formula for the computation of the 
electric field from the scalar and vector potentials is 
\begin{equation}
\vec E(\vec r, t) = 
- \vec\nabla \Phi(\vec r, t) - 
\frac{\partial}{\partial t} \vec A(\vec r, t) \,.
\end{equation}
In Coulomb gauge, since 
$\vec A_C(\vec r, t) = \vec A_{C\perp}(\vec r, t)$,
we can slightly rewrite this formula as 
\begin{equation}
\vec E(\vec r, t) = 
- \vec\nabla \Phi_C(\vec r, t) - 
\frac{\partial}{\partial t} \vec A_{C\perp}(\vec r, t) \,,
\end{equation}
and therefore 
\begin{equation}
\label{rr1}
\vec E_\parallel(\vec r, t) = 
- \vec\nabla \Phi_C(\vec r, t)  \,,
\qquad
\vec E_\perp(\vec r, t) = 
- \frac{\partial}{\partial t} \vec A_{C\perp}(\vec r, t) \,.
\end{equation}
These components fulfill, explicitly,
\begin{equation}
\label{rr2}
\vec\nabla \times \vec E_\parallel(\vec r, t) = 
- \left( \vec\nabla \times \vec\nabla \right) \, 
\Phi_C(\vec r, t) = 0 \,,
\end{equation}
and
\begin{equation}
\label{rr3}
\vec\nabla \cdot
\vec E_\perp(\vec r, t) = 
- \frac{\partial}{\partial t} 
\vec\nabla \cdot
\vec A_{C\perp}(\vec r, t) =0\,.
\end{equation}
Therefore, the full longitudinal component 
of the electric field is given by 
$\vec E_\parallel(\vec r, t) = - \vec\nabla \Phi_C(\vec r, t)$,
and the transverse component  in Coulomb gauge
is purely given by the time derivative of the transverse 
vector potential, without any ``admixture'' from the 
scalar potential.
In Coulomb gauge, the scalar potential $\Phi( \vec r, t)$
is given by an action-at-a-distance integral,
as a result of the instantaneous coupling of the 
potential to the source (charge density) according to Eq.~\eqref{rgauge1},
\begin{equation}
\label{phiCgen}
\Phi_C( \vec r, t) =
\frac{1}{4 \pi \epsilon_0} \,
\int \dd^3 r' \, \frac{1}{| \vec r - \vec r'|} \, 
\rho(\vec r', t) \,.
\end{equation}
The longitudinal component of the electric field is 
thus calculated as 
\begin{equation}
\label{ECparallel}
\vec E_\parallel(\vec r, t) = 
- \vec \nabla \Phi_C( \vec r, t) =
- \frac{1}{4 \pi \epsilon_0} \,
\vec\nabla \int \dd^3 r' \, \frac{1}{| \vec r - \vec r'|} \, 
\rho(\vec r', t) 
\end{equation}
which again is an action-at-a-distance integral.
The central result of the work of
Rohrlich~\cite{Ro2002} 
(the remarks of Jefimenko~\cite{Je2002comment},
Heras~\cite{He2003}
and of Rohrlich~\cite{Ro2002reply}
do not affect the derivation)
states that we can alternatively write the 
action-at-a-distance component of the 
electric field as a retarded integral,
\begin{equation}
\label{ECparallelGR}
\vec E_\parallel(\vec r, t) = 
- \int \dd^3 r' \, \int \dd t' \,
G_R(\vec r, t, \vec r', t') \, \left( \frac{1}{c^2} \, 
\frac{\partial}{\partial t'} \vec J_\parallel(\vec r', t') +
\vec\nabla' \rho(\vec r', t') \right) \,,
\end{equation}
where 
the retarded Green function $G_R(\vec r, t, \vec r', t')$
in coordinate-space is given in Eq.~\eqref{GR}.
The retarded Green function fulfills the 
defining equation
\begin{equation}\label{GRdiffeq}
\left( \frac{1}{c^{2}}\frac{\partial^2}{\partial t^2} 
- \vec\nabla^2 \right) G_R(\vec r, t, \vec r', t')  =
\frac{1}{\epsilon_0} \; 
\delta^{(3)}(\vec r - \vec r') \, \delta(t - t') \,,
\end{equation}
and its explicit form has been given in Eq.~\eqref{GR}.

Another perspective on this problem can be found 
by showing how the action-at-a-distance integral
for the longitudinal component of the electric 
field actually cancels when we add the time derivative of the 
Coulomb-gauge vector potential.
In order to find the solutions of 
Eqs.~\eqref{scalwave} and~\eqref{vectwave},
we use the retarded Green function given in Eq.~\eqref{GR}.
The solutions in Lorenz gauge can be written down immediately,
\begin{subequations}
\begin{align}
\label{phiL}
\Phi_L\left( \vec{r}, t\right) =& \;
\int \dd^3 r^{\prime } \dd t' \,
G_R(\vec r, t, \vec r', t') \,
\rho\left( \vec r', t' \right) \,,
\\[0.33ex]
\label{AL}
\vec A_L\left( \vec{r}, t\right) =& \;
\frac{1}{c^2} 
\int \dd^3 r^{\prime } \dd t' \,
G_R(\vec r, t, \vec r', t') \,
\vec J\left( \vec r', t' \right) \,.
\end{align}
\end{subequations}
The retarded Green function can be used in order to solve
Eq.~\eqref{rgauge2},  and we find the vector potential in 
Coulomb gauge,
\begin{align}
\label{ACLS}
\vec{A}_C( \vec r, t) =& \; \frac{1}{c^2} \int \dd^3 r' \int \dd t' \;
G_R ( \vec r, t, \vec r', t') \; \vec{J}_\perp ( \vec r' ,t') 
\nonumber\\[0.33ex]
=& \; \frac{1}{c^2} \int \dd^3 r' \int \dd t' \;
G_R ( \vec r, t, \vec r', t') \; \vec{J} ( \vec r' ,t') 
- \frac{1}{c^2} \int \dd^3 r' \int \dd t' \;
G_R ( \vec r, t, \vec r', t') \; \vec{J}_\parallel( \vec r' ,t') 
\nonumber\\[0.33ex]
=& \; \vec{A}_L( \vec r, t) + \vec{A}_S( \vec r, t) \,,
\end{align}
where we have used the identity 
$ \vec{J}_\perp ( \vec r' ,t')
= \vec{J} ( \vec r' ,t') - \vec{J}_\parallel( \vec r' ,t')$,
and $\vec{A}_L( \vec r, t)$ (Lorenz-gauge expression) and
$\vec{A}_S( \vec r, t)$ are defined in the obvious way.
The additional term $\vec{A}_S( \vec r, t)$ is relevant 
to the Coulomb gauge and reads
\begin{equation}
\label{AS}
 \vec{A}_S (\vec r,t)
= -\frac{1}{c^2} \int \dd^3 r' \int \dd t' \,
G_R ( \vec r, t, \vec r', t') \, \vec{J}_\parallel (\vec r',t') \,.
\end{equation}
The time derivative of $\vec{A}_S( \vec r, t)$
contributes to the electric field,
\begin{align}
E_S (\vec r,t) = -\frac{\partial}{\partial t} \vec{A}_S (\vec r,t)
&= \frac{1}{c^2} \int \dd^3 r' \int \dd t' 
\left[ \frac{\partial}{\partial t} \,
G_R ( \vec r, t, \vec r', t') \right] \, 
\vec{J}_\parallel (\vec r',t') \nonumber \\ 
&= -\frac{1}{c^2} \int \dd^3 r' \int \dd t' 
\left[ \frac{\partial}{\partial t'} \,
G_R ( \vec r, t, \vec r', t') \right] \, 
\vec{J}_\parallel (\vec r',t') \nonumber \\ 
&= \frac{1}{c^2} \int \dd^3 r' \int \dd t' 
 \, G_R ( \vec r, t, \vec r', t') \, 
\left[  \frac{\partial}{\partial t'} \vec{J}_\parallel (\vec r',t') 
\right] \,, 
\end{align}
where we have first transformed 
$\partial/\partial t \to - \partial/\partial t'$ and 
then used integration by parts to move the derivative 
on the current.
Using the identity in Eq.~\eqref{rgauge3}, this can be
rewritten in terms of the potential as
\begin{align}
E_S (\vec r,t)
&= \epsilon_0 \, \int \dd^3 r' \int \dd t' 
 \, G_R ( \vec r, t, \vec r', t') \, 
\vec\nabla'  \left[  \frac{1}{c^2}\frac{\partial^2}{\partial t^{' 2}} \,
\Phi_C\left( \vec{r}',t'\right)  
\right] \nonumber \\
&= \epsilon_0 \, \vec\nabla \int \dd^3 r' \int \dd t' 
 \, G_R ( \vec r, t, \vec r', t') \, 
\left[  \left( 
\frac{1}{c^2}\frac{\partial^2}{\partial t^{' 2}}
-\vec\nabla^{'\,2} \right) + \vec{\nabla}^{'\,2} \,
\right]  \Phi_C\left( \vec{r}',t'\right)  
\nonumber \\
&= \epsilon_0 \, \vec\nabla \int \dd^3 r' \int \dd t' 
 \, \left[  \left( 
\frac{1}{c^2}\frac{\partial^2}{\partial t^{2}}
-\vec\nabla^{2} \right) G_R ( \vec r, t, \vec r', t') \right] 
  \Phi_C\left( \vec{r}',t'\right)  
\nonumber \\
&\quad + \epsilon_0 \, \vec\nabla \int \dd^3 r' \int \dd t' 
 \, G_R ( \vec r, t, \vec r', t') \; \vec{\nabla}^{'2} 
  \Phi_C\left( \vec{r}',t'\right)  \,.
\end{align}
For the first term, we use partial integration twice,
and we also take advantage of the 
symmetry properties of the retarded Green function. With 
Eqs.~\eqref{rgauge1} and~\eqref{GRdiffeq}, we find
\begin{align}
\label{ES}
E_S (\vec r,t)
&= \epsilon_0 \vec\nabla \int \dd^3 r' \int \dd t' 
 \, \frac{1}{\epsilon_0} \delta^{(3)} \left(
 \vec r - \vec r' \right) \delta (t-t') \,
  \Phi_C\left( \vec{r}',t'\right)  
\nonumber \\
&\quad - \epsilon_0 \vec\nabla \int \dd^3 r' \int \dd t' 
 \, G_R ( \vec r, t, \vec r', t') \; \frac{1}{\epsilon_0} 
  \rho \left( \vec{r}',t'\right) \nonumber\\
&= \vec\nabla \Phi_C\left( \vec{r},t \right)  
 - \vec\nabla \int \dd^3 r' \int \dd t' 
 \, G_R ( \vec r, t, \vec r', t') \;  
  \rho \left( \vec{r}',t'\right) \nonumber\\ 
&=  \vec\nabla \Phi_C\left( \vec{r},t \right) 
- \vec\nabla \Phi_L\left( \vec{r},t \right)  \,.
\end{align}
This term cancels the longitudinal electric 
field in Coulomb gauge [$ = -\vec\nabla \Phi_C\left( \vec{r},t \right)$]
and adds the contribution to the electric 
field in Lorenz gauge, due to the 
gradient of the scalar potential [$ = -\vec\nabla \Phi_L\left( \vec{r},t \right)$].
In summary, the additional term 
$- \frac{\partial}{\partial t} \vec A_S(\vec r, t)$
due to the time derivative of the added
vector potential in Coulomb gauge cancels the 
gradient of the Coulomb gauge action-at-a-distance 
integral for the scalar potential and adds the 
Lorenz gauge gradient of the scalar potential.
Otherwise, this property is implied by the fact that $A_S$ is necessarily 
equal to $\vec\nabla \chi$,
where $\chi$ is the gauge potential from the transition from
Lorenz to Coulomb gauge. Here, we show the cancellation 
of the instantaneous interaction
by first coupling the potentials to the sources,
separately, in Lorenz and Coulomb gauge, using the 
retarded Green function, and then calculating the fields 
from the potentials. Our derivation is concise [Eqs.~\eqref{AS}---\eqref{ES}].

Comparing the formulas for the electric field 
in Coulomb and Lorenz gauge, the following identity follows immediately,
\begin{align}
\label{mainres}
\vec E_C(\vec r, t) =& \; -\vec\nabla \Phi_C(\vec r, t) 
- \frac{\partial}{\partial t} \vec A_C(\vec r, t) =
-\vec\nabla \Phi_C(\vec r, t) 
- \frac{\partial}{\partial t} \left( \vec A_L(\vec r, t) + \vec A_S(\vec r, t) \right) 
\nonumber\\[0.33ex]
=& \; 
-\vec\nabla \Phi_C(\vec r, t) 
- \frac{\partial}{\partial t} \vec A_L(\vec r, t) 
+ \left( \vec \nabla \Phi_C(\vec r, t)
- \vec \nabla \Phi_L(\vec r, t) \right)
=
- \vec \nabla \Phi_L(\vec r, t)
- \frac{\partial}{\partial t} \vec A_L(\vec r, t) 
= \vec E_L(\vec r, t) \,.
\end{align}
We have temporarily denoted the ``Coulomb gauge''
electric field as $\vec E_C(\vec r, t)$
and the ``Lorenz gauge''
electric field as $\vec E_L(\vec r, t)$,
even if both are actually equal due to gauge 
invariance, as shown.

Let us briefly summarize the findings. In the Coulomb gauge, the vector
potential is connected only to the transverse component of the current density.
We can write the transverse component of the current density as the difference
of the full current density, minus the longitudinal component of the current
density [Eq.~\eqref{ACLS}].  The full current density gives the same
contribution to the vector potential as in Lorenz gauge [Eq.~\eqref{phiL}].  In
Coulomb gauge, there is an additional term $\vec A_S$ in the vector potential
which corresponds to the negative of the longitudinal component of the current
density.  According to Eq.~\eqref{ES}, it can be written as a gradient vector
and therefore does not affect the result for the magnetic field,
which thus is the same as in Lorenz gauge.  The (negative of the) time
derivative of the additional term in the vector potential yields an additional
contribution to the electric field, in Coulomb gauge. The additional term in
the electric field can be transformed into two parts, the first of which
cancels the seemingly instantaneous electric field
contribution in Coulomb gauge, obtained from the Coulomb-gauge electric
potential, and the second yields the same result (the retarded one) as the
gradient of the electric potential $\Phi_L$ in Lorenz gauge.
In the end, the action-at-a-distance integral cancels, and 
the gauge invariance of the electric field is shown.

%
%
\section{Moving Charges in Different Gauges}
\label{pot}

%
%
\subsection{Moving Charges and Lorenz Gauge}

The representation~\eqref{GR} of the retarded Green 
function in coordinate space contains two 
Dirac-$\delta$ functions and is highly singular.
Therefore, it is certainly sensible to verify the 
conclusions reached thus far on the basis of 
a concrete problem where the actual form of the 
retarded Green function needs to be used
(Li\'{e}nard--Wiechert potentials for a 
moving point charge), because in addition
to the Green function, the charge and current 
distributions also become singular.
Indeed, the charge and current densities are 
\begin{subequations}
\label{movcharge}
\begin{align}
\label{movchargeRho}
\rho \left( \vec{r},t\right) =& \;
q \; \delta^{(3)}\left( \vec r- \vec R\left( t\right) \right) \,, \\
\label{movchargeJ}
\vec{J}\left( \vec r, t\right) =& \;
q \; \delta^{(3)}\left( \vec r- \vec R\left( t\right) \right) \;
\left[ \frac{\dd}{\dd t}\vec{R}\left( t\right) \right] \,,
\end{align}
\end{subequations}
where $\vec R(t)$ is the particle trajectory.
If we keep only the first term in square brackets in 
Eq.~\eqref{GR} and write 
\begin{equation}
\label{GRsimple}
G_R(\vec r, t, \vec r', t') \approx
\frac{c}{4\pi \, \epsilon_0} \; \frac{\Theta(t - t')}{|\vec r - \vec r'|} 
\delta\left( |\vec r - \vec r'| - c \, (t - t')\right) \,,
\end{equation}
then we can express the potentials of the 
moving charges in Eq.~\eqref{movcharge} as follows,
\begin{subequations}
\begin{align}
\Phi_L\left( \vec{r}, t\right) \approx & \;
\frac{q}{4 \pi \epsilon_0} \;
\int \dd^3 r^{\prime } \dd t' 
\left[ 
\delta^{(3)}\left( \vec r' - \vec R\left( t'\right) \right) 
\right] 
\left\{ \frac{\Theta \left( t-t' \right) }{| \vec r- \vec r' | }%
\delta \left( t - t' - \frac{| \vec r- \vec r' |}{c} \right) 
\right\} \,,\label{lorenzphi}
\\[0.33ex]
\vec A_L\left( \vec{r}, t\right) \approx & \;
\frac{q}{4 \pi \epsilon_0 c^2} \;
\int \dd^3 r^{\prime } \dd t' 
\left[ 
\delta^{(3)}\left( \vec r' - \vec R\left( t'\right) \right) \;
\dot{\vec R}\left( t' \right) \right] 
\left\{ \frac{\Theta \left( t-t' \right) }{| \vec r- \vec r' | }%
\delta \left( t - t' - \frac{| \vec r- \vec r' |}{c} \right) 
\right\} \,.\label{lorenzA}
\end{align}
\end{subequations}
We now carry out the integration over $\dd^3 r'$, 
which eliminates three integrations at once,
\begin{subequations}
\label{tleft}
\begin{align}
\Phi_L\left( \vec{r},t\right) =& \;
\frac{q}{4 \pi \epsilon_0} \;
\int \dd t' \, \Theta \left( t-t' \right) \,
\frac{1}{| \vec r - \vec R(t') | } \;
\delta \left( t' - \left( t - \frac{| \vec r- \vec R(t') |}{c} \right) \right) \,,
\\[1ex]
\vec{A}_L\left( \vec{r},t\right) =& \;
\frac{q}{4 \pi \epsilon_0 c^2} \;
\int \dd t' \, \Theta \left( t-t' \right) \,
\frac{\dot{\vec R}(t')}{| \vec r - \vec R(t') | } \;
\delta \left( t' - \left( t - \frac{| \vec r- \vec R(t') |}{c} \right) \right) \,.
\end{align}
\end{subequations}
The $\delta$ function peaks at
\begin{equation}
\label{tretarded}
t' = t_{\rm ret} = 
t - \frac{| \vec r- \vec R(t') |}{c} =
t - \frac{| \vec r- \vec R(t_{\rm ret}) |}{c} < t \,,
\end{equation}
or 
\begin{equation}
\label{condition}
t_{\rm ret} =
t - \frac{| \vec r- \vec R(t_{\rm ret}) |}{c} \,,
\qquad
c (t - t_{\rm ret})^2 = 
\left(  \vec r- \vec R(t_{\rm ret}) \right)^2 \,,
\end{equation}
so that the step function is always unity at the point where the 
Dirac-$\delta$ peaks. That means that all points on the trajectory of the 
particle for which the retardation condition is fulfilled,
contribute to the integrals.
Let us assume that the Dirac-$\delta$ peaks only once,
namely, at $t' = t_{\rm ret}$.
The integration over $\dd t'$ in Eq.~\eqref{tleft} still leads
to a nontrivial Jacobian as the argument of the 
Dirac-$\delta$ is nontrivial and needs to be differentiated
with respect to $t'$.
Indeed, we have
\begin{equation}
\label{deriv}
\frac{\dd}{\dd t'} \left( t' - t + \frac{| \vec r- \vec R(t') |}{c} \right) 
= 1 - \frac{1}{c} \, \frac{\dd  \vec R(t') }{\dd t'} \cdot
\frac{\vec r- \vec R(t')}{| \vec r- \vec R(t') |} \,,
\end{equation}
at $t' = t_{\rm ret}$. So,
the well-known Li\'{e}nard--Wiechert potentials in Lorenz gauge read
(see Ref.~\cite{Ja1998}),
\begin{subequations}
\begin{align}
\label{lienardphi}
\Phi_L\left( \vec{r},t\right) =& \;
\frac{q}{4 \pi \epsilon_0} \;
\frac{1}{| \vec r - \vec R(t_{\rm ret}) | } \;
\left( 1 - \frac{\dot{\vec R}(t_{\rm ret})}{c} \cdot 
\frac{\vec r- \vec R(t_{\rm ret})}{| \vec r- \vec R(t_{\rm ret}) |}
\right)^{-1} \,,
\\[0.33ex]
\label{lienardA}
\vec{A}_L\left( \vec{r},t\right) =& \;
\frac{q}{4 \pi \epsilon_0 c^2} \;
\frac{\dot{\vec R}(t_{\rm ret})}{| \vec r - \vec R(t_{\rm ret}) | } \;
\left( 1 - \frac{\dot{\vec R}(t_{\rm ret})}{c} \cdot 
\frac{\vec r- \vec R(t_{\rm ret})}{| \vec r- \vec R(t_{\rm ret}) |} \right)^{-1} \,.
\end{align}
\end{subequations}
Here, we write the equations in SI mksA units.
{\em The validity of this well-known derivation, which we 
recall for convenience, hinges upon the fact that no 
partial integrations involving the simplified Green function~\eqref{GRsimple} 
were necessary.}

%
%
\subsection{Coulomb Gauge and Current Components}

The charge and current density corresponding to the moving point charge are
given by Eq.~\eqref{movcharge}, and the Coulomb-gauge coupling to the 
potentials is given by Eq.~\eqref{rgauge}. 
The task is to calculate, explicitly, the
Li\'{e}nard--Wiechert potentials due to the 
moving charges, and to show the gauge invariance of the fields.
In particular, the question needs to be answered how the instantaneous
Coulomb interaction should be treated. 
The longitudinal component of the current is given by
\begin{align}
\label{Jparallel}
\vec{J}_\parallel\left( \vec{r},t\right) = & \;
\frac{q}{4\pi} \vec\nabla 
\frac{\partial}{\partial t}
\frac{1}{\left| \vec r- \vec R(t) \right| } \,.
\end{align}
It is advantageous to leave this result in symbolic form. We can 
easily verify that
$\vec{J}_\parallel$ carries the entire divergence of $\vec J$,
and that the curl of $\vec{J}_\parallel$ vanishes.
The transverse component of the point-particle 
current density can be written down immediately,
\begin{align}
\vec{J}_\perp\left( \vec{r},t\right) = & \;
\vec{J}\left( \vec{r},t\right) -
\vec{J}_\parallel\left( \vec{r},t\right) 
= q \; \dot{\vec R}(t) \, \delta^{(3)}\left( \vec r- \vec R\left( t\right) \right) 
- \frac{q}{4\pi} \vec\nabla 
\frac{\partial}{\partial t}
\frac{1}{\left| \vec r- \vec R(t) \right| } \,.
\nonumber
\end{align}
Its curl is given by
\begin{align}
& \vec\nabla \times \vec{J}_\perp\left( \vec{r},t\right) =
\vec\nabla \times 
\left( q \; \dot{\vec R}(t) \, 
\delta^{(3)}\left( \vec r- \vec R\left( t\right) \right) \right)
= -q \; \dot{\vec R}(t) \times \vec\nabla 
\delta^{(3)}\left( \vec r- \vec R\left( t\right) \right) 
=
\vec\nabla \times \vec{J}\left( \vec{r},t\right) \,.
\end{align}

%
%
\subsection{Scalar Potential in Coulomb Gauge}

The scalar potential in
Coulomb gauge is a solution of~\eqref{rgauge1}
and is easily calculated as the action-at-a-distance term
\begin{align}
\label{phiC1}
\Phi_C( \vec r, t) =& \; 
\frac{1}{4 \pi \epsilon_0} \, \frac{q}{| \vec r - \vec R(t)|} \,.
\end{align}
Let us try to verify the relation~\eqref{rgauge3},
which is easily done as follows,
\begin{align}
\epsilon_0\, \vec\nabla \left[ \frac{\partial }{\partial t}
\Phi_C\left( \vec{r},t\right) \right]
=& \; \epsilon_0\, \vec\nabla \frac{\partial }{\partial t}
\frac{1}{4 \pi \epsilon_0} \, \frac{q}{| \vec r - \vec R(t)|} 
= \frac{q}{4 \pi} \, \vec\nabla \frac{\partial }{\partial t}
\frac{1}{| \vec r - \vec R(t)|} 
= \vec{J}_\parallel(\vec r, t) \,.
\end{align}
At this point, the scalar potential in Coulomb gauge
is calculated as an action-at-a-distance integral.
It remains to investigate the vector potential 
in Coulomb gauge, which is responsible for the eventual 
cancellation of the action-at-a-distance solution~\eqref{phiC1}.
This is done in the next section.

%
%
\subsection{Vector Potential in Coulomb Gauge}

We recall that in Coulomb gauge, there is an 
additional term $\vec A_S$ in the vector potential 
given by Eq.~\eqref{AS}. In order to calculate it, 
we need the result~\eqref{Jparallel} for the 
longitudinal part of the current density. We write
\begin{align}
\label{coudiffA}
\vec{A}_S ( \vec r, t) 
&= -\frac{1}{c^2} \int \dd^3 r' \int \dd t' 
G_R ( \vec r, t, \vec r', t') \frac{q}{4 \pi} \,
\vec\nabla' \left( \frac{\partial }{\partial t'}
\frac{1}{| \vec r' - \vec R(t')|} \right)
\nonumber \\
& = -\frac{q}{(4 \pi)^2 \epsilon_0 c} 
\int \dd^3 r' \int \dd t' 
\left( \frac{\partial }{\partial t'} \left[ \vec\nabla'
\frac{1}{| \vec r' - \vec R(t')|} \right] \right) 
\frac{\Theta (t -t')}{| \vec r - \vec r' |} \\
& \qquad \times \left\{ \delta \left( |\vec r - \vec r' | - c ( t - t') \right) - 
\delta \left( |\vec r - \vec r' | + c ( t - t') \right) \right\} \,.
\nonumber
\end{align}
It now turns out to be very important to 
use the complete retarded Green function, when we perform
a partial integration with respect to $t'$
and shift the $t'$-derivative from the current term on the Dirac-$\delta$. 
When encountering a term from the $t'$-derivative acting on the 
Heavyside-$\Theta$, which gives $-\delta (t-t')$, then we shall 
find that it causes the two Dirac-$\delta$s from the Green 
function to have the same arguments and cancel. 
In conclusion, the only resulting term after the partial integration
with respect to $t'$ is 
\begin{equation}
\begin{split}
\vec{A}_S ( \vec r, t) & = 
\frac{q}{(4 \pi)^2\epsilon_0} \int \dd t' \int \dd^3 r' 
\left( \vec\nabla' \frac{1}{| \vec r' - \vec R(t')|} \right) \;
\frac{\Theta (t -t')}{| \vec r - \vec r' |} \\
& \qquad \times
\left\{ \delta' \left( |\vec r - \vec r' | - c ( t - t') \right) + 
\delta' \left( |\vec r - \vec r' | + c ( t - t') \right) \right\} \\
& = \frac{q}{(4 \pi)^2\epsilon_0} \int \dd t' \Theta (t -t')
\int \dd^3 \xi \left( \frac{\partial}{\partial \vec \xi}
\frac{1}{| \vec \xi - (\vec r - \vec R(t'))|} \right) \;
\frac{1}{| \vec \xi |} \;\\
& \qquad \times
\left\{ \delta' \left( |\vec \xi | - c ( t - t') \right) + 
\delta' \left( |\vec \xi | + c ( t - t') \right) \right\}  \,.
\end{split}
\end{equation}
The sign of the final result is clear from the fact that we 
differentiate the Dirac-$\delta$s with respect to $t'$, not $t$.
We have defined the new variable $\vec \xi = \vec r - \vec r'$, 
with $\dd \xi = - \dd r'$ and $\vec \nabla' = - \partial/\partial \vec \xi$.
Now, we can change the variable of
the gradient yet again to $\vec r$ and move it out of the integral.
We define 
\begin{equation}
\label{defx}
\vec x (t') = \vec r - \vec{R}(t')  
\end{equation}
and obtain
\begin{equation}
\vec{A}_S ( \vec r, t) = 
\frac{q}{(4 \pi)^2\epsilon_0} \vec\nabla \int \dd t' \Theta (t -t')
\int \dd^3 \xi \, \frac{1}{| \vec \xi - \vec x (t')|} \, \frac{1}{| \vec \xi |} 
\left\{ \delta' \left( |\vec \xi | - c ( t - t') \right) + 
\delta' \left( |\vec \xi | + c ( t - t') \right) \right\}  \,.
\end{equation}
Now, we expand the first of the terms into spherical harmonics,
using the well-known formula
\begin{equation}
\label{multipole}
\frac{1}{| \vec \xi - \vec x (t')|} = \sum_{\ell =0}^\infty \sum_{m=-\ell}^{\ell} 
\frac{4 \pi}{2 \ell +1} \frac{r_{<}^\ell}{r_{>}^{\ell+1}} 
Y_{\ell m} ( \Omega_\xi) \, Y_{\ell m}^* (\Omega_x) \,,
\qquad
r_< = \min(| \vec \xi |, x (t')) \,,
\quad
r_> = \max(| \vec \xi |, x (t')) \,,
\end{equation}
where $\Omega_\xi$ and $\Omega_x$ denote the solid angles for the 
vector variables $\vec\xi$ and $\vec x$, and obtain
\begin{equation}
\begin{split}
& \vec{A}_S ( \vec r, t) = \frac{q}{(4 \pi)^2\epsilon_0} 
\vec\nabla \int \dd t' \Theta (t -t')
\sum_{\ell =0}^\infty \sum_{m=-\ell}^{\ell} 
\int \dd \Omega_\xi \frac{4 \pi}{2 \ell +1} 
Y_{\ell m} ( \Omega_\xi) \, Y_{\ell m}^* (\Omega_x) 
\left[
\int \limits_{0}^{x(t')} \dd \xi \, 
\xi^2 \, \frac{1}{\xi} \frac{\xi^\ell}{x(t')^{\ell+1}}  
\right.
\\
& \times \left. \left\{ \delta' \left( \xi - c ( t - t') \right) + 
 \delta' \left( \xi + c ( t - t') \right) \right\} 
\!+ \!\!\! \int \limits_{x(t')}^{\infty} \!\!\! \dd \xi \,  
 \xi^2 \, \frac{1}{\xi} \frac{x(t')^\ell}{\xi^{\ell+1}}
 \left\{ \delta' \left( \xi - c ( t - t') \right) + 
 \delta' \left( \xi + c ( t - t') \right) \right\} \right] ,
\end{split}
\end{equation}
where we use $|\vec \xi| =\xi$ as well as $|\vec x(t')| = x(t')$,
and split the integral according to the 
$r_<$ versus $r_>$ symbols in Eq.~\eqref{multipole}.
The symbol $x(t') = | \vec x(t') | = | \vec r - \vec R(t) |$ is not being 
integrated over and therefore pulled out from the 
integrand into the prefactor.
Because $1/|\vec \xi|$ is a scalar, only the $\ell =0$, $m = 0$ 
component of the multipole expansion
contributes, and the expression simplifies to
\begin{equation}
\begin{split}
\vec{A}_S ( \vec r, t) &= \frac{q}{4 \pi\epsilon_0} \vec\nabla \int \dd t' \Theta (t -t')
 \left[  \frac{1}{x(t')}
 \int \limits_{0}^{x(t')} \dd \xi \, \xi \,  
 \left\{ \frac{\dd}{\dd \xi} \delta \left( \xi - c ( t - t') \right) + 
 \frac{\dd}{\dd \xi} \delta \left( \xi + c ( t - t') \right) \right\} \right. \\
 & \quad \left. + \int \limits_{x(t')}^{\infty} \dd \xi \,  
 \left\{ \frac{\dd}{\dd \xi} \delta \left( \xi - c ( t - t') \right) + 
 \frac{\dd}{\dd \xi} \delta \left( \xi + c ( t - t') \right) \right\} \right] \,.
\end{split}
\end{equation}
Here, we have written $\delta'$ as a derivative with respect to $\xi$.
The first term in square brackets is now integrated
by parts, whereas the last term can be integrated directly
because it constitutes a total differential. The result is
\begin{align}
\vec{A}_S ( \vec r, t) =& \;
\frac{q}{4 \pi \epsilon_0} \vec\nabla \int \dd t' \Theta (t -t')
\left\{ \left[  \frac{\xi}{x(t')}  
\left\{ \delta \left( \xi - c ( t - t') \right) + 
\delta \left( \xi + c ( t - t') \right) \right\} \right]_{0}^{x(t')}
\right. 
\\
& \; - \frac{1}{x(t')} \! \int \limits_{0}^{x(t')} \!\! \dd \xi \,  
\left\{ \delta \left( \xi - c ( t - t') \right) + 
\delta \left( \xi + c ( t - t') \right) \right\} \!+\! \left. 
\biggl[ \delta \left( \xi - c ( t - t') \right) + 
\delta \left( \xi + c ( t - t') \right) \biggr]_{x(t')}^{\infty}\!  \right\} .
\nonumber
\end{align}
After the cancellation of boundary terms (the limits must be considered
carefully), the expression reads 
\begin{equation}
\label{ACfordiv}
\begin{split}
\vec{A}_S ( \vec r, t) 
&= 
- \frac{q}{4 \pi \epsilon_0} \vec\nabla \int \dd t' \Theta (t -t') \,
\frac{1}{x(t')} \int \limits_{0}^{x(t')} \dd \xi \,   
 \left\{ \delta \left( \xi - c ( t - t') \right) + 
 \delta \left( \xi + c ( t - t') \right) \right\} \\
&= 
- \frac{q}{4 \pi \epsilon_0} \vec\nabla \int \limits^t_{-\infty} \dd t' \,
\frac{1}{|\vec r -\vec{R}(t')|} \int \limits_{0}^{|\vec r -\vec{R}(t')|} \dd \xi \,   
 \left\{ \delta \left( \xi - c ( t - t') \right) + 
 \delta \left( \xi + c ( t - t') \right) \right\} \,.
\end{split}
\end{equation}
We now use the same approximation~\eqref{GRsimple}
for the retarded Green function as in Lorenz gauge.
The second of the Dirac-$\delta$s only peaks for $t=t'$ within 
the $t'$ integration, and we can discard this contribution 
because the integration limits for the $t'$ integration ensure
that $t'$ cannot become larger than $t$.
On the other hand, if $t'$ is smaller than $t_{\rm ret} =
t - |\vec r -\vec{R}(t')|/c$, then the first Dirac-$\delta$ also cannot peak
because the $\xi$ integration stops at the upper limit of 
$\xi = |\vec r -\vec{R}(t')|$.
The net effect of the $\xi$ integration thus is that the 
$t'$ integration will be restricted to the interval~$t' \in (t_{\rm ret}, t)$,
\begin{equation}
\label{ALeibniz}
\vec{A}_S ( \vec r, t) 
= 
- \frac{q}{4 \pi \epsilon_0} \vec\nabla \int \limits^{t}_{t_{\rm ret}} \dd t'  \,
\frac{1}{|\vec r -\vec{R}(t')|} \,.
\end{equation}
We have thus obtained, in symbolic form, the vector potential 
for a moving particle in the Coulomb gauge.
We remember that according to Eq.~\eqref{ACLS},
the full Coulomb-gauge vector potential is given by
$\vec{A}_C( \vec r, t) = \vec{A}_L( \vec r, t) + \vec{A}_S( \vec r, t)$,
where $\vec A_L$ is given in Eq.~\eqref{AL} and for a moving point
particle in Eq.~\eqref{lienardA}. So, the Li\'enard-Wiechert potentials 
in Coulomb gauge are
\begin{subequations}
\label{potentialsC}
\begin{align}
\label{phiC}
\Phi_C \left( \vec r ,t\right) &= 
\underbrace{ \frac{q}{4 \pi \epsilon_0} 
\frac{1}{|\vec r -\vec{R}(t)|} }_{\mbox{action-at-a-distance}} \,, \\[0.33ex]
\label{AC}
\vec{A}_C ( \vec r, t) 
&= \frac{q}{4 \pi \epsilon_0 c^2} \;
\frac{\dot{\vec R}(t_{\rm ret})}{| \vec r - \vec R(t_{\rm ret}) | } \;
 \left( 1 - \frac{\dot{\vec R}(t_{\rm ret})}{c} \cdot 
\frac{\vec r- \vec R(t_{\rm ret})}{| \vec r- \vec R(t_{\rm ret}) |} \right)^{-1}
- \underbrace{
\frac{q}{4 \pi \epsilon_0} \vec\nabla \int \limits^{t}_{t_{\rm ret}} \dd t'  \,
\frac{1}{|\vec r -\vec{R}(t')|} }_{\mbox{action-at-a-distance}}\,.
\end{align}
\end{subequations}
To the best of our knowledge, these compact integral 
representations have not yet appeared 
in the literature for the Li\'{e}nard--Wiechert 
potentials in Coulomb gauge. These equations illustrate 
the instantaneous character of the scalar potential
but also the existence of an additional instantaneous term
in the vector potential, whose calculation necessitates 
the knowledge of the trajectory of the particle over the 
time interval $t' \in (t_{\rm ret}, t)$.
The latter term is responsible
for the required cancellation of the non-causal terms in the calculation
of the observable electric and magnetic fields.

The general cancellation mechanism given by Eq.~\eqref{ES}
can be verified as follows.
We recall that $t_{\rm{ret}}$ is an implicit function of the form
\begin{equation}
F(t,t_{\rm{ret}}) =  t - t_{\rm{ret}} -\frac{\vec r -\vec{R}(t_{\rm{ret}})}{c} = 0 \,.
\end{equation}
The implicit function theorem then states that the
derivative of $t_{\rm{ret}}$ with respect to $t$ is
\begin{equation}
\frac{\partial t_{\rm{ret}}}{\partial t} = - \left( 
\frac{\partial F(t,t_{\rm{ret}})}{\partial t_{\rm{ret}}} \right)^{-1}
\frac{\partial F(t,t_{\rm{ret}})}{\partial t} = \left( 1 - \frac{\dot{\vec R}(t_{\rm ret})}{c} \cdot 
\frac{\vec r- \vec R(t_{\rm ret})}{| \vec r- \vec R(t_{\rm ret}) |} \right)^{-1}\,,
\end{equation}
a result which can alternatively
be derived according to Eq.~\eqref{res_time} below. 
The derivative of $\vec{A}_S$ with respect to $t$ is thus found as
\begin{equation}
\begin{split}
- \frac{\partial}{\partial t} \vec{A}_S ( \vec r, t) 
&= 
\frac{q}{4 \pi \epsilon_0} \vec\nabla \frac{\partial}{\partial t}
 \int \limits^{t}_{t_{\rm ret}} \dd t'  \,
\frac{1}{|\vec r -\vec{R}(t')|} \\
&= 
\frac{q}{4 \pi \epsilon_0} \vec\nabla \frac{1}{|\vec r -\vec{R}(t)|} 
- \frac{q}{4 \pi \epsilon_0} \vec\nabla 
\frac{1}{|\vec r -\vec{R}(t_{\rm{ret}})|} 
\frac{\partial t_{\rm{ret}}}{\partial t} \\
&= 
\frac{q}{4 \pi \epsilon_0} \vec\nabla \frac{1}{|\vec r -\vec{R}(t)|} 
- \frac{q}{4 \pi \epsilon_0} \vec\nabla \left[
\frac{1}{|\vec r -\vec{R}(t_{\rm{ret}})|} 
\left( 1 - \frac{\dot{\vec R}(t_{\rm ret})}{c} \cdot 
\frac{\vec r- \vec R(t_{\rm ret})}{| \vec r- \vec R(t_{\rm ret}) |} \right)^{-1}  \right] \\
&= \vec \nabla \Phi_C(\vec r, t)  - \vec \nabla \Phi_L(\vec r, t) \,.
\end{split}
\end{equation}
In Appendix~\ref{appa}, we verify that the 
result given in Eq.~\eqref{potentialsC} is consistent 
with the general formula given in Eq.~(3.10) of Ref.~\cite{Ja2002}
for potentials in Coulomb gauge. 
While our result could be derived from Eq.~(3.10) of Ref.~\cite{Ja2002},
this alternative derivation is not immediate;
we present a detailed derivation in Eq.~\eqref{jacksonC}
below which illustrates
the mechanism by which the action of the $\vec\nabla$ operator on the 
action-at-a-distance 
integral in Eq.~\eqref{AC} generates nontrivial boundary terms
from the differentiation of the lower integration limit $t' = t_{\rm ret}$.
Our treatment using the full retarded Green function simplifies the 
inclusion of the nontrivial Jacobian from the time integration.

%
%
\section{Conclusions}
\label{conclu}

The main theme of the current investigation is to show that 
the action-at-a-distance solution~\eqref{phiC} 
of the Coulomb gauge scalar potential coupled to the 
charge density, does not cause any problems 
with respect to the principle of causality.
In order to provide for a concrete 
framework for the calculations, we consider the 
scalar and vector potentials corresponding to 
a moving point particle given in Eqs.~\eqref{movchargeRho}
and~\eqref{movchargeJ}.

According to Eq.~\eqref{rgauge2}, the vector potential in Coulomb gauge is
transverse and it couples only to the transverse component of the current
density.  Furthermore, according to Eq.~\eqref{ACLS}, the Coulomb gauge vector
potential can be written as the difference of the full Lorenz gauge vector
potential, minus the vector potential generated by the longitudinal part of the
current density.  The time derivative of the latter component of the vector
potential, generated by the longitudinal component of the current density,
yields an additional term for the electric field, which supplements the
gradient of the (instantaneous) action-at-a-distance integral from the charge
density.  We show for the general case as well as by an explicit calculation for 
the moving point charge that the time derivative of the
additional term in the vector potential in Coulomb gauge can be written as the
sum of two terms. The first term cancels the action-at-a-distance integral,
while the second is equal to the gradient of the Lorenz gauge scalar
potential. This is summarized in Eq.~\eqref{mainres}.

Thus, the electric field in Coulomb gauge is shown to be equal to the electric
field in Lorenz gauge, as it should be. Moreover, our calculation provides for
an additional perspective on the problem: namely, the instantaneous integral
over the charge density, whose gradient contributes to the electric field in
Coulomb gauge, actually cancels against the additional time derivative from the
supplement to the vector potential in Coulomb gauge.

In addition to Rohrlich's arguments~\cite{Ro2002}, we are able to say that the
instantaneous action-at-a-distance integral for the scalar potential cancels
against a term generated by the time derivative of the vector potential in
Coulomb gauge and thus {\em does not contribute} to the physically observable
electric field. The conclusions of Ref.~\cite{Ro2002} and of the current work
can thus be summarized as follows. The action-at-a-distance
integral~\eqref{phiCgen} for the Coulomb gauge scalar potential can (i) be
rewritten in terms of a retarded integral, over different source terms, which
happens to be equal to the instantaneous interaction integral by virtue of the
particular properties of the Coulomb gauge, and (ii) does not contribute to the
electric field because its gradient is precisely canceled by an additional
contribution to the time derivative of the vector potential present only in
Coulomb gauge.

Compact integral expressions for the Coulomb-gauge potentials 
generated by a moving charge 
(the ``Coulomb-gauge Li\'{e}nard--Wiechert potentials'') 
are written down in Eq.~\eqref{potentialsC}.
They are consistent with the general treatment of Ref.~\cite{Ja2002}
and generalize earlier work on this 
problem~\cite{La1999coulomb,Hn2004,Es2007},
where results were obtained for specific kinematic conditions
(e.g., uniform motion).

%
%
\section*{Acknowledgments}

The authors acknowledge helpful conversations with J.~Sly.
This work was supported by the NSF and by the 
National Institute of Standards and Technology
(precision measurement grant).

\appendix

%
%
\section{Verification of the Li\'{e}nard--Wiechert Potential in Coulomb Gauge}
\label{appa}

We compare to the treatment outlined in Ref.~\cite{Ja2002}.
First, we need a few nontrivial derivatives.
The retarded time fulfills the equation
\begin{equation}
\label{deft_ret}
t_{\rm ret} = t - \frac{| \vec r - \vec R(t_{\rm ret}) |}{c} \,.
\end{equation}
Under the variation $t \to \delta t$ and 
$t_{\rm ret} \to t_{\rm ret} + \delta t_{\rm ret}$, 
keeping Eq.~\eqref{deft_ret} fulfilled,
we have
\begin{align}
t_{\rm ret} + \delta t_{\rm ret}
=& \; t + \delta t - \frac{| \vec r - 
\vec R(t_{\rm ret}+ \delta t_{\rm ret})|}{c} 
= t + \delta t + \delta t_{\rm ret} \, \dot{\vec R}(t_{\rm ret}) \cdot 
\vec\nabla \frac{| \vec r - \vec R(t_{\rm ret})|}{c} 
\,.
\end{align}
Solving for $t_{\rm ret}$, we find
\begin{equation}
\label{res_time}
\frac{\partial t_{\rm ret}}{\partial t} =
\frac{\delta t_{\rm ret}}{\delta t} =
\left( 1 - \dot{\vec R}(t_{\rm ret}) \cdot 
\vec\nabla \frac{| \vec r - \vec R(t_{\rm ret})|}{c} \right)^{-1} \,.
\end{equation}
By contrast, varying 
$\vec r \to \vec r + \delta \vec r$ and
$t_{\rm ret} \to t_{\rm ret} + \delta t_{\rm ret}$,
we obtain
\begin{align}
t_{\rm ret} + \delta t_{\rm ret}
=& \; t - \frac{| \vec r + \delta \vec r - 
\vec R(t_{\rm ret}+ \delta t_{\rm ret})|}{c} 
\nonumber\\[0.33ex]
=& \; t - \delta \vec r \cdot 
\vec\nabla \frac{| \vec r - \vec R(t_{\rm ret})|}{c} 
+ \delta t_{\rm ret} \, 
\dot{\vec R}(t_{\rm ret}) 
\cdot 
\vec\nabla \frac{| \vec r - \vec R(t_{\rm ret})|}{c} 
\,,
\end{align}
and thus
\begin{equation}
\label{res_grad}
\vec\nabla t_{\rm ret} = \frac{\delta t_{\rm ret}}{\delta \vec r} 
= - \frac{\vec r - \vec R(t_{\rm ret})}{c \, | \vec r - \vec R(t_{\rm ret})|} \;
\left( 1 - \dot{\vec R}(t_{\rm ret}) \cdot 
\vec\nabla \frac{| \vec r - \vec R(t_{\rm ret})|}{c} 
\right)^{-1} \,.
\end{equation}
The retarded character of the interaction is still visible.
We now refer to Ref.~\cite{Ja2002}.
The variable $R = | \vec x - \vec x' |$ is defined
in the text after Eq.~(2.5) of Ref.~\cite{Ja2002}
and we use this identification of $\vec x$ in the following
instead of the definition given in Eq.~\eqref{defx}.
The time $t'$ in Jackson's formulas is
always [also see text after Eq.~(2.5) of Ref.~\cite{Ja2002}]
\begin{equation}
t_{\rm ret} = t' = t - R/c = 
t - |\vec x - \vec x' |/c = 
t - |\vec x - \vec R(t_{\rm ret}) |/c \,,
\end{equation}
and thus defined by an implicit equation.
Then, according to Eq.~(3.10) of Ref.~\cite{Ja2002},
\begin{align}
\label{ACJACKSON}
4 \pi \epsilon_0 c \vec A_C(\vec r, t)
=& \; \frac{1}{c} \int \dd^3 x' \,
\left( \frac{1}{R} \left[ \vec J(\vec x', t') - c 
\hat{R} \rho(\vec x', t') \right]_{t' = t_{\rm ret}} \right) 
+
c \, \int \dd^3 x' \, \frac{\hat{R}}{R^2} \, 
\int_0^{R/c} \dd \tau \, \rho(\vec x', t - \tau)
\\[0.33ex]
=& \; \frac{1}{c} \int \dd^3 r' \,
\frac{1}{| \vec r - \vec r' |} \, \vec J(\vec r', t_{\rm ret})  
- \int \dd^3 r' \, 
\frac{\vec r - \vec r'}{| \vec r - \vec r'|^2} \,
\rho(\vec r', t_{\rm ret}) 
+ c \, \int_{t_{\rm ret}}^{t} \dd \tau \, 
\int \dd^3 r' \, \frac{\vec r - \vec r'}{|\vec r - \vec r'|^3} \, 
\rho(\vec r', \tau) \,,
\nonumber
\end{align}
where $4\pi\epsilon_0 c$ is restored in the SI mksA system we use here.
The integral~\eqref{ACJACKSON} can immediately be applied to 
non-singular charge and current distributions.
However, for a point charge, we have to plug the 
singular expressions for the charge and current 
densities~\eqref{movcharge} into Eq.~\eqref{ACJACKSON}, so that we cannot 
carry out the $\dd^3 r'$ integrations immediately because 
$t_{\rm ret}$ itself is a function of $\vec r'$. 
It is convenient to re-introduce the $t'$ integrations, 
then carry out the $\dd^3 r'$ integrations, 
and then, carry out the $t'$ integrations which lead to the 
Jacobians. The derivation is rather straightforward but 
a detailed discussion necessitates a number of intermediate steps,
which we would like to present in detail,
\begin{align}
\label{jacksonC}
4 \pi \epsilon_0 c \, \vec A_C(\vec r, t)
=& \; \frac{1}{c} 
\int \dd^3 r' \,
\frac{1}{| \vec r - \vec r' |} \, q \; 
\left. \frac{\partial \vec R(t')}{\partial t'} \right|_{t = t_{\rm ret}} \;
\delta^{(3)}\left( \vec r'- \vec R\left( t_{\rm ret} \right) \right) 
- \int \dd^3 r' \, \frac{\vec r - \vec r'}{| \vec r - \vec r'|^2} \,
q \; \delta^{(3)}\left( \vec r- \vec R\left( t_{\rm ret}\right) \right) 
\nonumber\\[0.33ex]
& \; + c \, \int_{t_{\rm ret}}^{t} \dd \tau \, 
\int \dd^3 r' \, \frac{\vec r - \vec r'}{|\vec r - \vec r'|^3} 
q \; \delta^{(3)}\left( \vec r' - \vec R\left( t_{\rm ret} \right) \right) 
\nonumber\\[0.33ex]
=& \; \frac{1}{c} \int \dd^3 r' \, \int \dd t' \,
\frac{1}{| \vec r - \vec r' |} \, q \; 
\frac{\partial \vec R(t')}{\partial t'} \;
\delta^{(3)}\left( \vec r'- \vec R\left( t' \right) \right) \,
\delta\left( t' - t + \frac{|\vec r - \vec r' |}{c} \right) 
\nonumber\\[0.33ex]
& \;
- \int \dd^3 r' \,  \int \dd t' \,
\frac{\vec r - \vec r'}{| \vec r - \vec r'|^2} \,
q \; \delta^{(3)}\left( \vec r- \vec R\left( t' \right) \right) \,
\delta\left( t' - t + \frac{|\vec r - \vec r' |}{c} \right) 
\nonumber\\[0.33ex]
& \; + q \, c \, \int_{t_{\rm ret}}^{t} \dd \tau \, 
\int \dd^3 r' \, \frac{\vec r - \vec r'}{|\vec r - \vec r'|^3} 
\delta^{(3)}\left( \vec r' - \vec R\left( t_{\rm ret} \right) \right) 
\nonumber\\[0.33ex]
=& \; \frac{q}{c} \int \dd t' \,
\frac{1}{| \vec r - \vec R(t') |} \, 
\frac{\partial \vec R(t')}{\partial t'} \;
\delta\left( t' - t + \frac{|\vec r - \vec R(t') |}{c} \right) 
\nonumber\\[0.33ex]
& \;
- q \, \int \dd t' \,
\frac{\vec r - \vec R(t') }{| \vec r - \vec R(t') |^2} \,
\delta\left( t' - t + \frac{|\vec r - \vec R(t') |}{c} \right) 
+ q \, c \, \int_{t_{\rm ret}}^{t} \dd \tau \, 
\frac{\vec r - \vec R(\tau)}{|\vec r - \vec R(\tau)|^3} 
\nonumber\\[0.33ex]
=& \; \frac{q}{c}
\frac{\dot{\vec R}(t_{\rm ret})}{| \vec r - \vec R(t_{\rm ret}) |} \,
\left( 1 - \frac{\dot{\vec R}(t_{\rm ret})}{c} \cdot 
\frac{\vec r- \vec R(t_{\rm ret})}{| \vec r- \vec R(t_{\rm ret}) |} \right)^{-1}
\nonumber\\[0.33ex]
& \; - q \; \frac{\vec r - \vec R(t_{\rm ret})}{| \vec r - \vec R(t_{\rm ret}) |^2} \,
\left( 1 - \frac{\dot{\vec R}(t_{\rm ret})}{c} \cdot 
\frac{\vec r- \vec R(t_{\rm ret})}{| \vec r- \vec R(t_{\rm ret}) |} \right)^{-1}
+ q \, c \, \int_{t_{\rm ret}}^{t} \dd \tau \, 
\frac{\vec r - \vec R(\tau)}{|\vec r - \vec R(\tau)|^3} 
\nonumber\\[0.33ex]
=& \; \frac{q}{c}
\frac{\dot{\vec R}(t_{\rm ret})}{| \vec r - \vec R(t_{\rm ret}) |} \, 
\frac{\partial t_{\rm ret}}{\partial t} \;
+ q \, c \; \frac{1}{| \vec r - \vec R(t_{\rm ret}) |} \,
\vec\nabla t_{\rm ret}
- q \, c \, \int_{t_{\rm ret}}^{t} \dd \tau \, 
\vec\nabla \frac{1}{|\vec r - \vec R(\tau)|} 
\nonumber\\[0.33ex]
=& \; \frac{q}{c}
\frac{\dot{\vec R}(t_{\rm ret})}{| \vec r - \vec R(t_{\rm ret}) |} \,
\left( 1 - \frac{\dot{\vec R}(t_{\rm ret})}{c} \cdot 
\frac{\vec r- \vec R(t_{\rm ret})}{| \vec r- \vec R(t_{\rm ret}) |} \right)^{-1}
- q \, c \; \vec\nabla \int_{t_{\rm ret}}^{t} \dd \tau \, 
\frac{1}{|\vec r - \vec R(\tau)|} \,.
\end{align}
The expression in the last line is finally equal to the 
result given in Eq.~\eqref{potentialsC}.
The results~\eqref{deriv},~\eqref{res_time} and~\eqref{res_grad} 
have been used at various places during the derivation.


\begin{thebibliography}{10}
\bibitem{LorenzLorentz}
Biographies of Ludvig Lorenz and Hendrik Antoon Lorentz are freely available
  from various online encyclopediae.
\end{thebibliography}
\end{document}